# Volviendo a las raíces: cómo crear fotos estroboscópicas a partir de vídeos digitales

# Going back to the roots: how to create stroboscopic photos from digital videos

**Daniel Baccino[a], Álvaro Suárez[b], Arturo C. Martí[c],**

[a] Consejo de Formación en Educación, dbaccisi@gmail.com
[b] Consejo de Formación en Educación, alsua@outlook.com
[c] Facultad de Ciencias, Universidad de la República, marti@fisica.edu.uy

**Resumen:**

Mostramos aquí cómo crear a partir de filmaciones digitales usando el conocido software Tracker fotos estroboscópicas con el objetivo de analizar distintos tipos de movimientos. La ventaja de este procedimiento es que es posible analizar la foto impresa o en la pantalla de forma intuitiva para los estudiantes. Luego de presentar una perspectiva histórica de la utilización de las fotos estroboscópicas en la enseñanza media discutimos varios ejemplos: el movimiento de un autito a control remoto, un choque elástico en el plano, el movimiento de un proyectil y también un experimento de electromagnetismo, más concretamente, la descarga de un capacitor medida utilizando un multímetro analógico.

**Abstract:**

We show here how to create from digital films using the well-known software Tracker stroboscopic photos in order to analyze different types of movements. The advantage of this procedure is that it is possible to analyze the printed photo or on a computer screen in an intuitive way for the students. After presenting a historical perspective of the use of stroboscopic photos in secondary education we discuss several examples: the movement of a remote control car, an elastic planar collision, the movement of a projectile and also an experiment in electromagnetism, more specifically, the discharge of a capacitor measured using an analog multimeter.

## 1. Introducción

Desde hace más de cuatro décadas el uso de fotografías estroboscópicas es un recurso muy extendido en la enseñanza a nivel medio, especialmente en temas de cinemática y dinámica. La fotografías de exposición múltiple o estroboscópicas consisten en la toma, mediante una cámara convencional con el obturador abierto, del movimiento de un objeto iluminado por una luz estroboscópica, es decir, un flash disparado a intervalos regulares. Este tipo de fotografía tiene la ventaja que resulta bastante claro e intuitivo para los estudiantes entender cualitativamente las características de un movimiento. Además, una vez impresa, es posible determinar distintas características de los movimientos, tales como desplazamientos, velocidades o aceleraciones medias y poner de manifiesto el carácter vectorial de estas magnitudes o discutir sus incertidumbres.

Estas fotografías son un recurso muy versátil para analizar diversos fenómenos, en una o dos dimensiones, como el movimiento de caída libre, la conservación de la cantidad de movimiento lineal en el choque de dos cuerpos, el movimiento circular de un objeto sobre una superficie con baja fricción o el movimiento relativo, entre otros.

Antes del advenimiento de las herramientas digitales, la realización de fotos estroboscópicas presentaba una serie de dificultades. En efecto, uno de los inconvenientes era poder contar con un cámara fotográfica que permitiera tomar fotografías en modo "B" (obturador abierto mientras se mantiene presionado el pulsador) o "T" (el obturador se abre el presionar una vez el pulsador y se cierra al presionar por segunda vez). Otro elemento necesario era el flash estroboscópico y todo el montaje debía ubicarse en un laboratorio que pudiera oscurecerse razonablemente. Por otro lado, el costo de los insumos, películas, revelado y ampliación también era un elemento a tener muy en cuenta.

Una forma de contrarrestar estas dificultades era usar fotografías estroboscópicas obtenidas por terceros, ya sea de otros grupos de años anteriores o de textos de Física, tales como el PSSC [1]. En particular, las fotografías de dicho texto fueron empleadas por generaciones de estudiantes, y se destacan por varios aspectos. En la sección siguiente se presenta una perspectiva histórica del uso de las fotos estroboscópicas.

Con el tiempo fueron surgiendo algunas alternativas o mejoras ingeniosas como la que se muestra en la figura 1. En este caso se utiliza una cámara convencional pero no requiere de un flash estroboscópico. Con el obturador de la cámara fotográfica abierto, se registraba el movimiento de un cuerpo que tenía montado un dispositivo emisor de luz intermitente (por ejemplo un pequeño LED). En dicha figura se muestra una fotografía obtenida por este procedimiento e indicamos la ubicación del emisor de luz con una flecha sobre la imagen original. Estas fotografías fueron tomadas en la década del 80 por el profesor Leonardo De Simone.

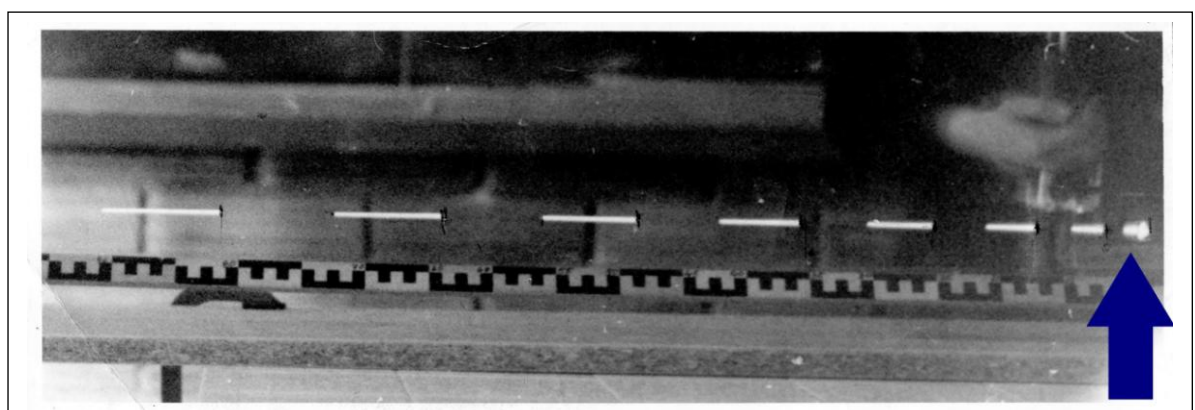

Figura 1. Foto estroboscópica utilizada para analizar movimiento rectilíneo acelerado. Autor: Leonardo De Simone, aproximadamente 1980.

La técnica de las fotos estroboscópicas también ha dado lugar a otras derivaciones. Una de ellas muy relacionada conceptualmente es la conocida práctica de la cinta y el *timer*. En este caso el movimiento se registra gracias a timer (habitualmente un timbre desarmado, o las nuevas versiones "electrónicas") cuyo percutor golpea un papel carbónico y una cinta que a su vez está unida a un objeto en movimiento. Esta técnica presenta varias desventajas. Por un lado, es poco precisa y requiere sujetar el objeto a estudiar a una cinta y por otra las marcas se hacen en un punto fijo, lo que conlleva una dificultad conceptual para el estudiante. Otro aspecto que merece destacarse es el caso del texto de Hecht [2] que utiliza profusamente dibujos o representaciones de sistemas físicos evolucionando en el tiempo de la forma en que lo haría una foto estroboscópica. En este caso, si bien no son imágenes fotográficas superpuestas, se trata de diagramas que simbolizan la misma idea.

En este artículo exponemos el procedimiento para obtener fotos estroboscópicas a partir de vídeos digitales usando el software Tracker [3,4]. Los vídeos pueden ser obtenidos por los propios estudiantes utilizando teléfonos inteligentes o cámara digitales. Se discuten también distintas actividades que se pueden realizar utilizando esta técnica y se muestran algunos resultados.

## 2.- El uso de fotos estroboscópicas en algunos libros

Mediante un relevamiento no exhaustivo de algunos libros disponibles en laboratorios de física del CES y del IPA, identificamos en ellos la cantidad de páginas que contenían fotos estroboscópicas. Se han contado las imágenes cuyos autores las denominan fotos estroboscópicas[1] en el cuerpo del libro.

Los gráficos de la figura 2 muestran los resultados en dos formatos. En la parte superior se presenta un gráfico en el que se indica la cantidad de fotos estroboscópicas en función del año de edición en español. En la parte inferior de la figura se muestra la misma variable en función del apellido del primer autor del libro. En ambos gráficos se destaca que, de los libros indagados, la edición del PSSC (1962) contiene la mayor cantidad (21) de fotos estroboscópicas, alejándose notoriamente de la media (cercana a 5) del conjunto analizado.

---

[1]O sinónimos: foto de exposición múltiple, de múltiples destellos, de flash múltiple, entre otros.

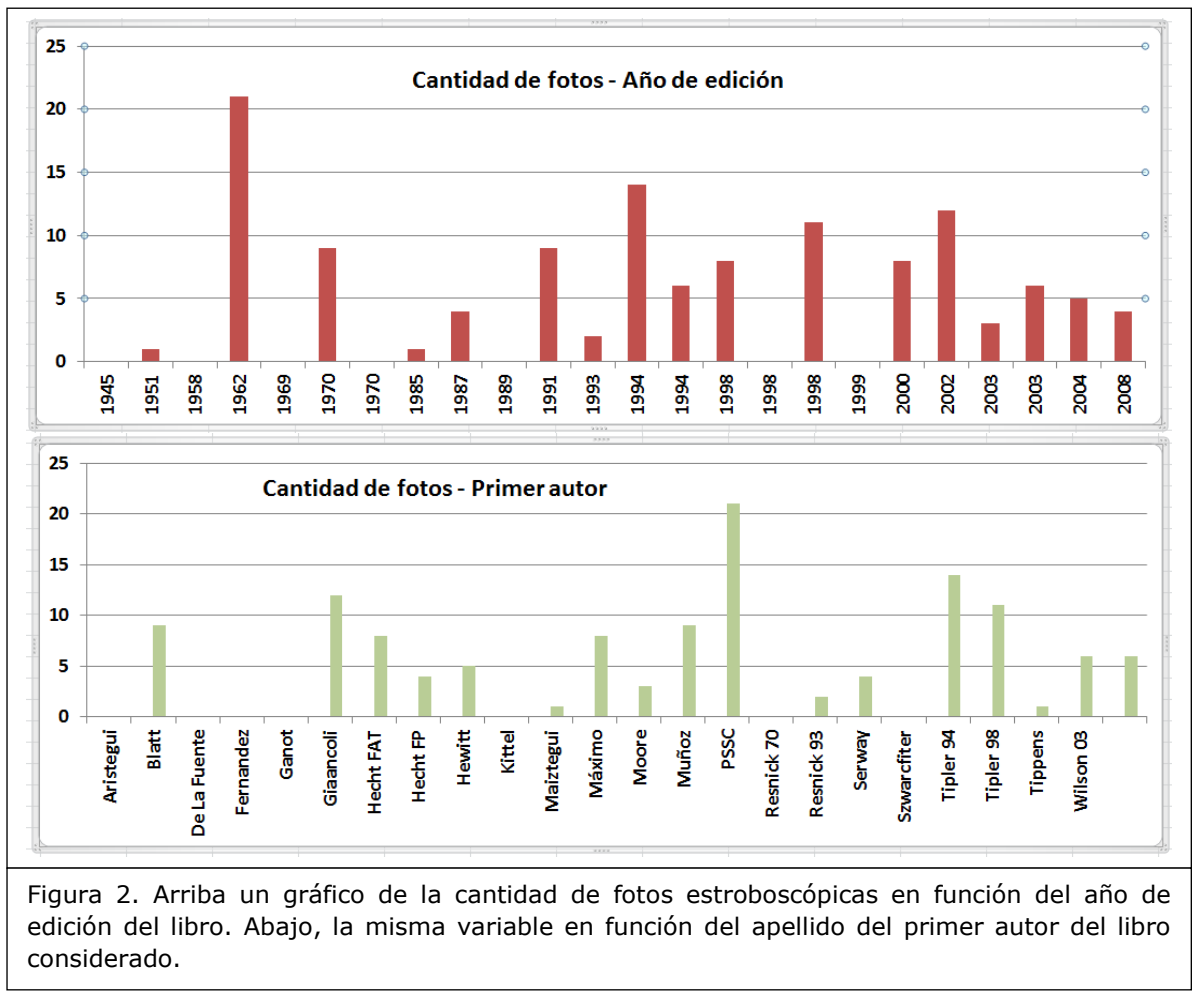

Figura 2. Arriba un gráfico de la cantidad de fotos estroboscópicas en función del año de edición del libro. Abajo, la misma variable en función del apellido del primer autor del libro considerado.

Del análisis cualitativo del material resulta que varias de las imágenes originales de PSSC están presentes en libros editados posteriormente. En ese sentido este libro resulta en un fuerte promotor del uso de la herramienta y fuente de fotografías, que otros autores incluyeron en sus obras.

Ese análisis muestra también una característica de las fotos estroboscópicas incluidas en esta edición del PSSC: la mayoría de ellas son imágenes destacadas en el libro y tienen la información necesaria para realizar análisis cuantitativos (frecuencia del estroboscopio, escala definida, masas de los cuerpos cuando es relevante). Este elemento distingue a este libro, ya que en la amplia mayoría de los restantes las fotografías no muestran esa información, y en algunos casos no se destacan como en aquel.

## 3. Fotografías estroboscópica a partir de videos digitales

En la actualidad es posible obtener fotografías estroboscópicas a partir de vídeos digitales de una forma muy rápida y utilizarlas para analizar sistemas físicos [5-7]. Si se tienen ciertos cuidados en la obtención del vídeo, se puede generar una imagen final de excelente calidad para analizar un movimiento, con la ventaja que el estudiante experimenta con el movimiento en el laboratorio, en el corredor, en el patio o en cualquier otro lugar. Esta posibilidad permite tener un producto final que proviene de la experiencia directa del estudiante favoreciendo la contextualización, en contraste con la antigua práctica de trabajar sobre una fotocopia de una foto estroboscópica proveniente de un libro.

Dentro del abanico de posibilidades que hay para generar fotos estroboscópicas con herramientas digitales, en este artículo nos centraremos en el popular paquete de programas gratuito Tracker creado por Douglas Brown para analizar videos [3,4]. Un tutorial de carácter general sobre el uso de Tracker se puede consultar en [8]. Dentro de los usos menos conocidos de esta herramienta, nos centraremos ahora en la posibilidad de obtener imágenes mediante la superposición de cuadros de un vídeo digital de manera sencilla e intuitiva.

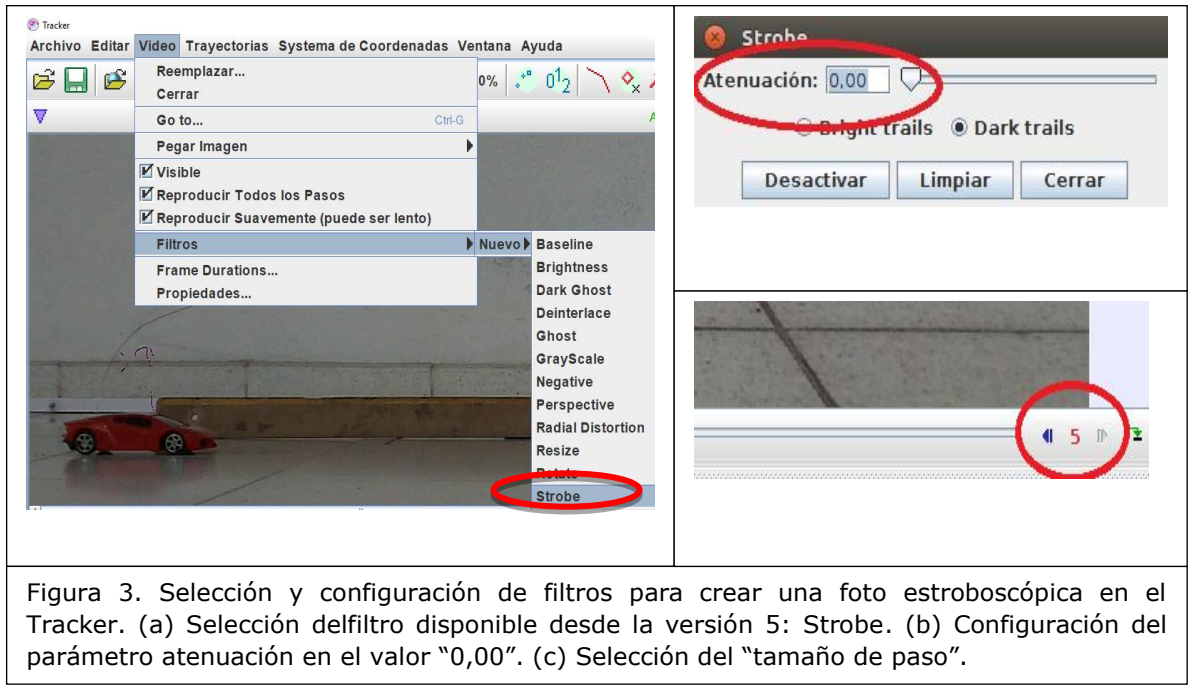

Figura 3. Selección y configuración de filtros para crear una foto estroboscópica en el Tracker. (a) Selección delfiltro disponible desde la versión 5: Strobe. (b) Configuración del parámetro atenuación en el valor "0,00". (c) Selección del "tamaño de paso".

Luego de ingresar al software Tracker, cargamos el video de interés (*Archivo → Abrir*… desde la barra de herramientas). Para generar la foto estroboscópica se requiere aplicar un filtro digital al video, al que se accede por el trayecto: Video → *Filtros → Nuevo*, como muestra la figura 3(a)*.* Al menos desde la versión 5 de Tracker se dispone del filtro *Strobe*[2] diseñado especialmente para nuestro fin.Al seleccionar este filtro aparecen dos opciones de configuración: la selección entre *Dark trials* o *Bright trials* se hace en función del brillo relativo entre el objeto cuyo movimiento se quiere seguir y el fondo[3]; para el parámetro *Atenuación* es conveniente seleccionar el valor 0,00 (figura 3b). Dependiendo de las características particulares del video y la situación filmada, puede ser conveniente seleccionar un tamaño de paso adecuado (figura 3c), para evitar tener una superposición de cuadros muy densa. Se corre el video y el software va superponiendo los cuadros en una sola imagen. El resultado final se exporta como imagen en formatos estándar. Un resultado obtenido al procesar el video de un auto a control remoto se expone en la figura 4. Para acceder a una guía "paso a paso" para realizar una foto estroboscópica, consulte la referencia [9].

---

[2]En versiones anteriores la herramienta para crear estas imágenes es el filtro *Dark Ghost*, y su configuración es similar a la reseñada para *Strobe*.
[3]Cuando el objeto en movimiento es más oscuro que el fondo, se debe seleccionar la opción Dark trials, en caso contrario, se elige la opción Bright trials.

## 4. Algunos resultados

En el primer ejemplo se estudia el movimiento unidimensional de un auto a control remoto. En la figura 4 se ve un ejemplo de fotografía estroboscópica. La escala de longitud puede ser obtenida fácilmente a partir de las medidas de las baldosas, claramente visibles en la foto. A partir de la foto estroboscópica se puede determinar la posición del auto para distintos instantes de tiempo, lo que permite construir un gráfico x(t) y a partir del mismo determinar por ejemplo la velocidad del auto en diferentes instantes. El auto a control remoto tiene la particularidad respecto a otros posibles movimientos que se pueden estudiar en enseñanza media, que alcanza una velocidad máxima en pocos segundos, siendo por ende un movimiento cuya aceleración no es constante.

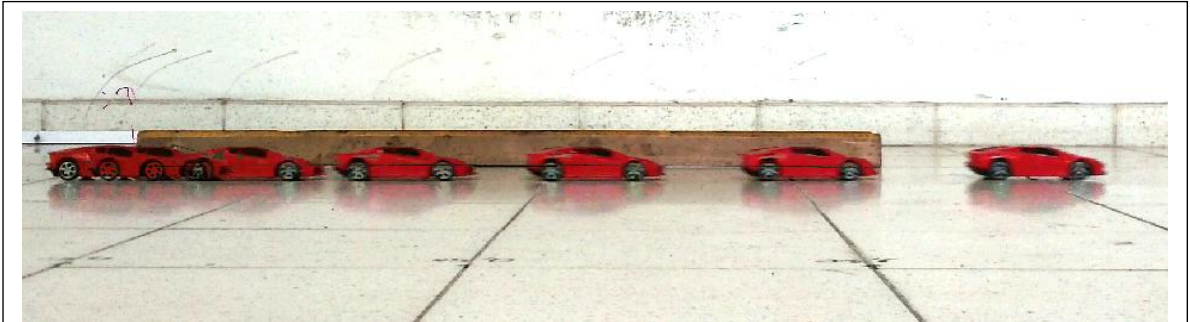

Figura 4. Un auto a control remoto se mueve en un trayecto rectilíneo en un corredor del IPA. Laboratorio de Física del IPA, 2018.

En la figura 5 se muestra la aplicación de la técnica a una interacción en dos dimensiones, más concretamente, el choque de fichas sobre un juego de tejo que opera como una mesa sin rozamiento. Este montaje apunta a estudiar problemas de conservación de cantidad de movimiento.

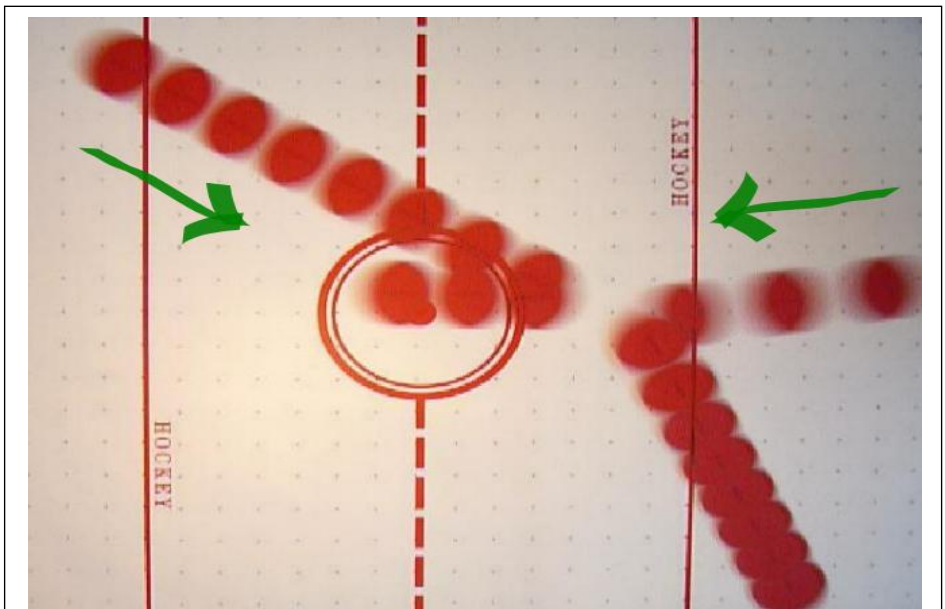

Figura 5. Fotografía estroboscópica de un choque en dos dimensiones sobre una mesa de tejo en un plano cenital. Dos fichas, ambas inicialmente en movimiento, chocan cerca del centro de la mesa. Laboratorio de Física del IPA, 2008.

En la figura 6 se muestra el resultado de aplicar el filtro Strobe a una secuencia de imágenes tomadas del artículo de Trinidad [10]. En dicho artículo el autor presenta la filmación digital como una herramienta para estudiar una caída, con algunas herramientas disponibles en ese contexto histórico.

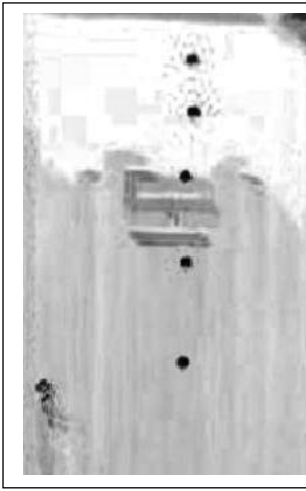

Figura 6. La imagen de la figura muestra el resultado de aplicar el filtro Strobe de Tracker a una secuencia de imágenes. Dado que no fue posible encontrar el video original, a partir del que se exponen las imágenes en el artículo, se procedió de la siguiente manera para obtener la imagen de esta figura: se cargó en el Tracker, en lugar del video una secuencia numerada de imágenes; con el objetivo de mejorar el resultado final se aplicó el filtro Negative; y finalmente el filtro Strobe para generar la imagen estroboscópica. Observe que la salida de la vertical del cuerpo está asociada, seguramente, al procesamiento manual de las imágenes originales.

En la figura 7 se muestra la foto estroboscópica obtenida de analizar la filmación de una esfera realizando un movimiento de proyectiles.

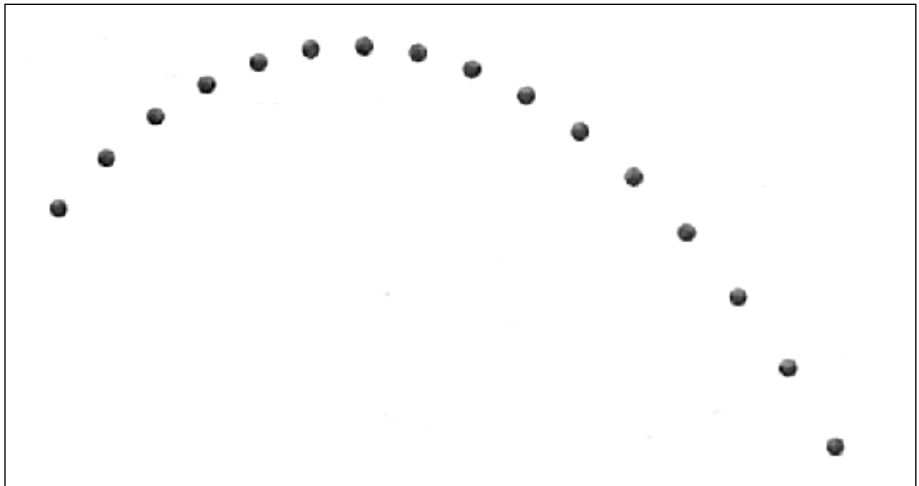

Figura 7. Fotografía estroboscópica de un movimiento de dos dimensiones. Laboratorio de Física del IPA, 2018.

Finalmente, en la figura 8 presentamos una foto de exposición múltiple de un multímetro analógico conectado a un capacitor electrolítico durante una descarga.

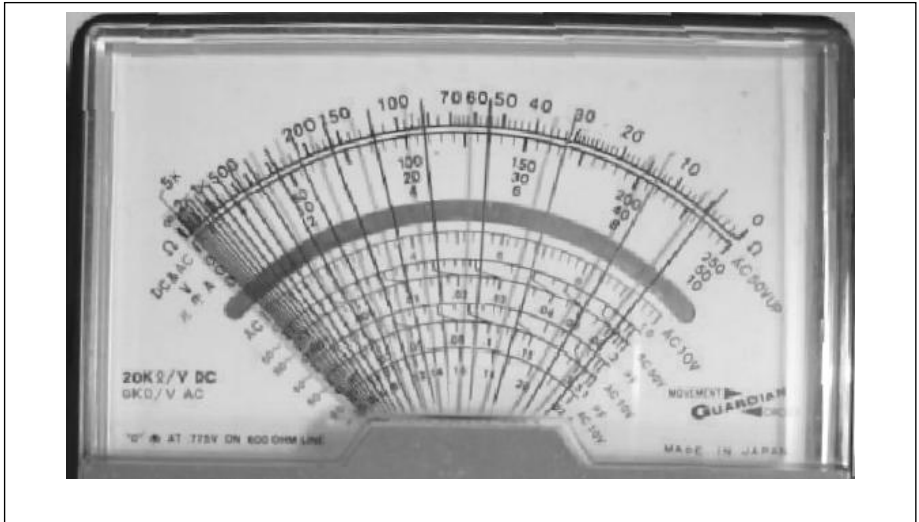

Figura 8. Fotografía estroboscópica de la aguja de un multímetro analógico mientras mide la diferencia de potencial en bornes de un capacitor durante una descarga.

## 5. Comentarios finales

En este trabajo presentamos una técnica para crear fotos estroboscópicas a partir de un vídeo digital. Consideramos firmemente que la herramienta presentada tiene un alto valor pedagógico y didáctico, permitiendo a los estudiantes poder participar en todos los aspectos de la construcción de la foto estroboscópica, desde el diseño experimental, la filmación del fenómeno físico involucrado y confección de la misma.

La sencillez y rapidez para obtener la foto estroboscópica es una de sus grandes virtudes, no siendo necesario un conocimiento previo del sofware Tracker, ya que las mismas se construyen con un par de simples pasos. Como consecuencia de esto y considerando que los videos pueden ser creados por los estudiantes con sus dispositivos móviles, pueden incluso realizar las fotos estroboscópicas fuera del aula. Esto permite además que los alumnos puedan trabajar desde su casa, a su propio ritmo, independizándose del laboratorio en el caso que sea necesario. De esta manera se incluye otro posible uso de los Smartphones para interactuar con el mundo físico fuera del aula más allá de los conocidos sensores [11-13].

Los ejemplos presentados son simplemente algunos de los posibles experimentos para los que se pueden construir fotos estroboscópicas [4-6]. Otras posibles actividades son el movimiento circular -pudiéndose construir una foto simplemente marcando un punto del sistema giratorio-, el análisis de la propagación de un pulso de onda, el movimiento de un péndulo.

Por último cabe señalar que aunque el software Tracker permite realizar un seguimiento automático del movimiento de un cuerpo, trazando en tiempo real los gráficos de posición y velocidad en función del tiempo, el uso de las fotos estroboscópicas sigue teniendo un valor en sí mismo. Trabajando con las mismas el estudiante pone en juego conceptos tales como posición, desplazamiento, velocidad y aceleración, midiendo dichas cantidades directa o indirectamente a partir del análisis de las fotos.